\documentclass{webofc}
\usepackage[varg]{txfonts}   
\wocname{New Frontiers in Physics 2014}
\woctitle{3rd International Conference on New Frontiers in Physics}
\begin{document}
\title{Measuring Antimatter Gravity with Muonium\\[-.75in]
\rightline{\normalsize\rm IIT-CAPP-14-03}\\[.5in]}
%
%

\author{Daniel M. Kaplan\inst{1}\fnsep\thanks{\email{kaplan@iit.edu}} \and
        Klaus Kirch\inst{2,3}
        \and
        Derrick Mancini\inst{4}
        \and
        James D. Phillips\inst{5}
         \and
        Thomas J. Phillips\inst{1}
         \and
        Thomas J. Roberts\inst{1,6}
         \and
        Jeff Terry\inst{1}
}

\institute{Illinois Institute of Technology, Chicago, Illinois 60616, USA 
\and
           Paul Scherrer Institute, Villigen, Switzerland
\and
           ETH, Z\"{u}rich, Switzerland
\and
           Argonne National Laboratory, Argonne, Illinois 60439, USA           
\and
           Harvard Smithsonian Center for Astrophysics, Cambridge, Massachusetts 02138, USA
\and
           also at Muons, Inc., Batavia, Illinois 60510, USA
          }

\abstract{%
 
The gravitational acceleration of antimatter, $\bar{g}$, has never been directly measured and could bear importantly on our understanding of gravity, the possible existence of a fifth force, and the nature and early history of the universe. Only two avenues for such a measurement appear to be feasible: antihydrogen and muonium. The muonium measurement requires a novel, monoenergetic, low-velocity, horizontal muonium beam directed at an atom interferometer. The precision three-grating interferometer can be produced in silicon nitride or ultrananocrystalline diamond using state-of-the-art nanofabrication. The required precision alignment and calibration at the picometer level also appear to be feasible. With 100 nm grating pitch, a 10\% measurement of $\bar{g}$ can be made using some months of surface-muon beam time, and a 1\% or better measurement with a correspondingly larger exposure. This could constitute the first gravitational measurement of leptonic matter, of 2nd-generation matter and, possibly, the first measurement of the gravitational acceleration of antimatter.
}
\maketitle
\section{Introduction}
\label{intro}
Indirect tests imply stringent limits on the gravitational acceleration of antimatter~\cite{Alves}:
$\bar{g} / g -1 < 10^{-7}$. Such limits are inferred based on the varying amounts of virtual antimatter presumed to constitute a portion of nuclear binding energies in various elements. (It is of course unclear to what extent these limits apply to muonium, as virtual antimuons surely play a negligible role in the nucleus.)
Of the attempts at a direct test, none has yet achieved significance: the only published direct limit to date, on antihydrogen~\cite{Amole}, is
$-65 < \bar{g} / g < 110$.

Besides antihydrogen, only one other experimental approach appears practical: measurements on muonium (M), a $\mu^+ e^-$ hydrogenic atom.
We are developing a precision 3-grating muonium atom-beam interferometer to measure $\bar{g}$. Such a measurement will constitute a unique test of the gravitational interaction of leptonic and 2nd-generation matter with the gravitational field of the earth.

\section{Experimental Approach}
The key issue that must be addressed in a muonium experiment is the short lifetime of the muon, $\tau=2.2\,\mu s$. Thus in one lifetime, assuming the ``null hypothesis'' of standard gravity, an M atom has a gravitational deflection of only 
\begin{equation*}
\frac{1}{2}g\tau^2=24\,{\rm pm}\,.
\end{equation*}
Measuring so small a deflection would be a substantial challenge even using state-of-the-art diffraction gratings and alignment and stabilization methods. This leads us to focus on measurements using ``old'' muons.
It is straightforward to show that the statistical optimum is obtained with a measurement time of four lifetimes, for which the deflection is 16 times greater:
\begin{equation*}
\frac{1}{2}g(4\tau)^2=379\,{\rm pm}\,.
\end{equation*}
Whether statistics will dominate such a measurement is as yet unclear. A common-sense guesstimate is that, on the contrary, systematics will likely play a significant role. Thus, e.g.,  a measurement time of six lifetimes,  statistically somewhat better than  one lifetime, may prove to be optimal, giving a null-hypothesis deflection of 850\,pm.

\subsection{Interferometer}
\begin{figure}
\centering
\includegraphics[width=13cm,clip]{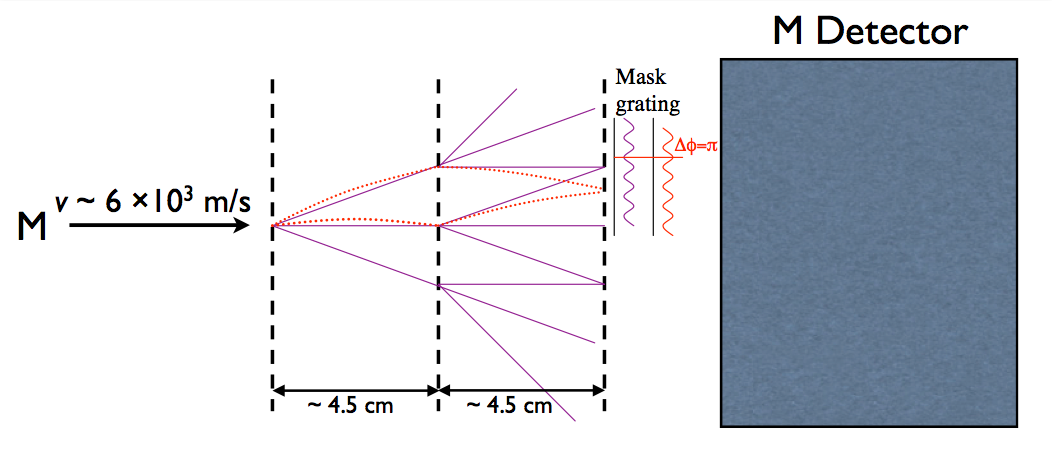}
\caption{Schematic diagram of measurement concept, with monoenergetic muonium (M) beam, incident from left, impinging on 3-grating Mach-Zehnder interferometer followed by muonium detector employing a coincidence measurement technique. Since the muonium beam undergoes a vertical gravitational acceleration, so does the interference pattern.}
\label{fig-MZ}       
\end{figure}

We propose to measure the gravitational deflection by means of a three-grating Mach--Zehnder  interferometer~\cite{Phillips} (Fig.~\ref{fig-MZ}). In this, the de Broglie wave corresponding to an incident M atom interferes with itself, and the interference pattern created by the first two gratings is sampled by translating the third grating up and down by means of piezoelectric actuators. The result is a sinusoidally varying modulation of the counting rate, with period equal to the grating pitch. The gravitational deflection causes a phase offset of this interference pattern with respect to that of a straight-through path.

\begin{figure}
\centering
\includegraphics[width=10cm,clip]{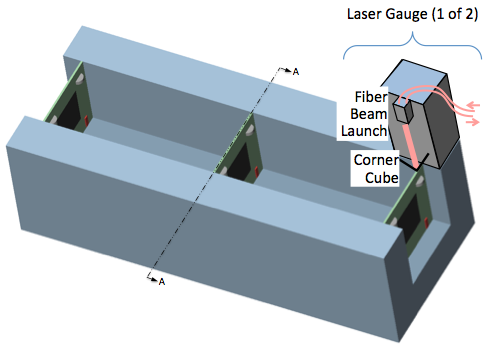}
\caption{Schematic diagram of mounting concept, with PDH-locked laser gauge indicated at one grating corner}
\label{fig-PDH}       
\end{figure}

Calibrating the  straight-through phase to the required accuracy by direct measurement of the relative grating offsets would be challenging. We therefore propose to measure it concurrently with the muonium phase measurement using a soft X-ray source of wavelength comparable to that of the M beam. Another challenge is maintaining the desired alignment to a small fraction of 100\,pm. 
This can be done by means of Pound--Drever--Hall (PDH)-locked laser tracking frequency gauges~\cite{Phillips-OL} mounted on at least one corner of each of two gratings (Fig.~\ref{fig-PDH}). The required dimensional stability can be achieved using Si${_3}$N${_4}$-film gratings mounted to a single-crystal silicon optical bench, operated cryogenically in order to take advantage of silicon's nearly zero cryogenic temperature coefficient of expansion. (An alternate grating material, ultrananocrystalline diamond, is also under consideration.)
Plausible state-of-the-art target grating parameters are 100\,nm pitch and 1\,cm$^2$ area.

\subsection{Beam}
\begin{figure}
\centering
\includegraphics[width=10cm,clip]{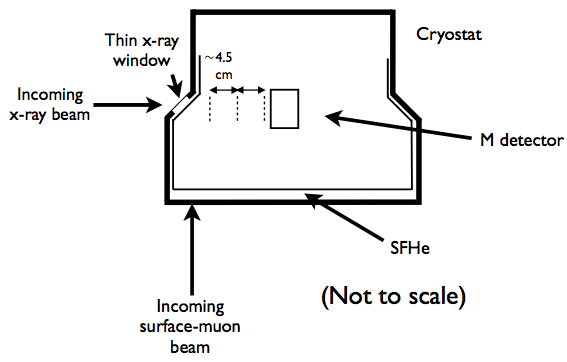}
\caption{Experiment concept sketch}
\label{fig-cryo}       
\end{figure}

It has been proposed to use a novel M beam produced by stopping decelerated surface muons in a thin layer of superfluid helium (SFHe)~\cite{Kirch}. Such a beam is anticipated to be monoenergetic (a requirement in order not to wash out the M gravitational phase difference) due to the immiscibility of hydrogen in SFHe. Taking into account the mass-dependent chemical potential of hydrogenic atoms in SFHe, the M beam is expected to be expelled perpendicular to the SFHe surface at a speed of 6,300\,m/s, with energy spread $\Delta E/E=0.2$\%. In order to obtain a horizontal M beam, a reflective SFHe surface at 45$^\circ$ is employed, as shown in Fig.~\ref{fig-cryo}, taking advantage of the well-known tendency of SFHe to flow up walls and coat the inside of its container. Note that the exact horizontality of the beam is not crucial, thanks to the incident-angle independence of the Mach--Zehnder arrangement~\cite{Chang-etal}.

\subsection{Sensitivity}
A sensitivity estimate has been given by Kirch~\cite{Kirch}:
\begin{equation*}
\Delta \bar{g}=0.3 g\, /\sqrt{\rm \#\,days} \,,
\end{equation*}
assuming an incident M rate of $10^5$/s, as estimated for the upgraded beam proposed at the Paul Scherrer Institute. At this rate the determination of the sign of $\bar{g}$ could be accomplished with minimal ($\sim$\,1\,day) running time (once the beam and  apparatus are installed, shaken down, and calibrated), or in proportionately longer running time with a beam of lower intensity. A first measurement will suffice to determine whether we live in a so-called ``Dirac--Milne'' universe (one in which $\bar{g}=-g$)~\cite{Dirac-Milne}. More challenging and time-consuming measurements could detect whether $\bar{g}/g=1\pm\epsilon$, for $\epsilon\sim10^{-2}$ or less.

\section{Theoretical Considerations}

We here briefly recap ideas developed more fully by Nieto and Goldman~\cite{Nieto-Goldman}. As is well known, experimentally, gravity is consistent with being a pure tensor interaction. However, in the most general quantum field theories there should also be scalar and tensor terms. In the absence of a successful quantum field theory of gravity we are left in a somewhat speculative realm, but it seems not unreasonable to attribute the lack of experimental evidence for non-tensor behavior to a near-exact cancelation between the scalar and vector terms.

The vector terms are of opposite signs for matter--matter and matter--antimatter interactions. A good cancelation in the former case would then allow quite a large deviation between $\bar{g}$  and $g$. Whether this means $\epsilon\sim10^{-2}$, $10^{-10}$, something in between, or something very much smaller, is difficult to predict based on current knowledge. This would imply an experiment-driven regime for the present\,---\,as has been the case, e.g., for searches for $\mu\to e\gamma$ decays until recent years.
More generally such a search has the potential to uncover an elusive fifth force, such as might be expected to couple to lepton- or lepton-generation number~\cite{Lee-Yang}.
To take the first experimental step into such an intriguing {\it terra incognita} would be a rare privilege indeed!

\section{Cosmological Considerations}
\label{sec-1}
Theories (see, e.g., \cite{Dirac-Milne}) in which antimatter repels matter (so-called ``antigravity'') offer simple explanations of a number of key cosmological puzzles:
(1) the cosmic baryon asymmetry; (2) galactic rotation curves and the binding of galaxy clusters; (3) the apparent cosmic acceleration; and (4) the ``horizon'' and ``flatness'' problems. The solutions of each of these problems in the antigravity scenario are briefly summarized in the following subsections.

\subsection{Cosmic baryon asymmetry}
Self-gravitating but mutually repulsive clusters of matter and antimatter form randomly interspersed matter and antimatter domains~\cite{Dirac-Milne}.
Thus there is no baryon asymmetry. The repulsion of matter and antimatter domains over billions of years ensures negligible production of annihilation radiation.

\subsection{Anomalous gravitational clustering}
In an antigravity theory there must be virtual gravitational dipoles (matter--antimatter pairs). Unlike the electromagnetic case, these are repulsive, giving ``anti-shielding,'' and strengthening the force of gravity at large distances~\cite{Hajdukovic-DM}.
Thus there is no need for dark matter.

\subsection{Cosmic acceleration}
Interspersed, mutually repulsive matter and antimatter clusters
counteract the  gravitational deceleration of the universe's expansion, leading to a constant rate of recession~\cite{Dirac-Milne}. Although differing from the usual, accelerating universe, interpretation, this is statistically consistent with the supernova data.
Thus there is no need for dark energy.

\subsection{Horizon and flatness problems}
Slower expansion of the early universe means that all parts are causally connected.
Thus there is no need for inflation~\cite{Dirac-Milne}.

\subsection{Occam's razor}
It is intriguing to note that the standard $\Lambda$CDM cosmology requires four postulates, for each of which there is little  or no independent corroboration: (1) anomalous {\em CP} violation far in excess of the experimentally observed Standard Model level; (2) dark matter; (3) dark energy; and (4) inflation. What would Occam say?

\section{Conclusion}
An experiment to measure the gravitational acceleration of muonium atoms is currently in an early R\&D stage. If the effort comes to fruition and the measurement is made, the consequences of a ``non-standard'' result could be immense. Even the null result would be ``one for the textbooks.'' A thorough account of the current knowledge of gravity would  include the disclaimer that it has never been directly tested with antimatter. If the experiment herein discussed succeeds, future textbooks will be able to say that experimental tests have demonstrated the applicability of the Principle of Equivalence even to antimatter\,---\,or not!

\section*{Acknowledgments}
We have benefited from discussions with the late Andy Sessler, and from the support of the IPRO program~\cite{IPRO} at Illinois Institute of Technology. The  development of a suitable muon and muonium beam is supported by the Swiss National Science Foundation, grant \#200020\_146902.

\end{document}